\documentstyle[twoside,fleqn,espcrc2,fps]{article}

\newcommand{\vep}{\varepsilon}

\makeatletter
\def\sitemize{\ifnum \@itemdepth >3 \@toodeep\else \advance\@itemdepth \@ne
\edef\@itemitem{labelitem\romannumeral\the\@itemdepth}%
\list{\csname\@itemitem\endcsname}{%
  \topsep 0pt plus 1pt \parsep 0pt plus 1pt \itemsep\parsep
  \def\makelabel##1{\hss\llap{##1}}}\fi}

\makeatother

\title{Spectrum and scaling in a strongly coupled fermion-gauge-scalar
  model}

\author{Wolfgang Franzki\address{Institute of Theoretical Physics E, RWTH
    Aachen, D-52056 Aachen, Germany}\thanks{Speaker.}
        and
        Xiang-Qian Luo\address{HLRZ c/o Forschungszentrum KFA, D-52425 J{\"u}lich, Germany \\
          and Deutsches Elektronen-Synchrotron DESY, D-22603 Hamburg, Germany}        
        \thanks{Work done in collaboration with C. Frick and J. Jersak.}%
}

\begin{document}

\begin{abstract}
  The strongly coupled lattice gauge models show an interesting mechanism of
  dynamical mass generation.  If a suitable continuum limit can be found, one
  may think of it as an alternative to the Higgs mechanism.  We present data
  on the spectrum, obtained in the model with U(1) gauge symmetry with
  dynamical fermions.  They indicate that the fermion mass scales in the
  vicinity of the whole chiral phase transition line.  In contrast to this,
  the composite scalar boson mass seems to get small only in the region near
  the endpoint E of the Higgs phase transition.  Thus this point is the most
  interesting candidate for approaching the continuum limit. The masses of
  fermion--antifermion bound states are also discussed.
\end{abstract}

\maketitle

\section{A Fermion-Gauge-Scalar model}
It seems to be a generic feature of strongly coupled gauge-theories to give
fermions a mass due to dynamical chiral symmetry breaking. Some models use
this fact to explain a heavy fermion mass generation (see~\cite{FrJe95b} for
references).  In our current work, we try to consider such a possibility by
examining a model with U(1) gauge symmetry.

We use three fields in our HMC simulation:
\begin{sitemize}
\item a compact U(1) symmetric gauge-field $U_{x,\mu}$ with gauge coupling
  $\beta=\frac{1}{g^2}$,
\item a charged staggered fermion field $\chi_x$, describing 4 fermions in the
  continuum and
\item a charged complex scalar field $\phi_x$ with fixed length and hopping
  parameter $\kappa$.
\end{sitemize}
For technical reasons we introduce a bare fermion mass $m_0$, although the
model is meant in the chiral limit $m_0 = 0$. The precise action and a
schematic phase diagram can be found in~\cite{FrJe95b}. All calculations
presented in this paper are obtained with dynamical fermions.

As described in some detail in~\cite{LuFr95}, this model has for strong gauge
coupling a chiral-phase-transition ($\chi$PT) line. Part of this line also
includes a Higgs phase transition and is of first order. On the left of the
endpoint {\bf E} of this Higgs phase transition (at small $\beta$) the line is
of second order ({\bf NE} line).  The chiral condensate
vanishes on the whole line.

In this paper we want to discuss the spectrum. In the 2$^{\rm nd}$ section we will
explain the basic properties of the spectrum. Some more details can be found
in~\cite{FrJe95a}. In the 3$^{\rm rd}$ section we then show our first numerical
results. More results will be included in a forthcoming paper~\cite{FrFr95}.

\section{Spectrum}
\subsection{Fermions}
Because we consider the model in the confinement region, only neutral
particles survive in the spectrum. To shield the charge we construct a
composite fermion ${\rm F}=\phi^+\chi$.  Due to the strong gauge coupling
it is strongly bounded.  Here one sees the important role of the scalar field.
It protects the fermions from getting confined.  Think of those fermions being
a heavy quark (top).

\subsection{Mesons}
As in QCD there is a large number of fermion--antifermion bound states which
we call by analogy mesons.  Up to now we only looked for those described by
local operators, which are the first four entries in the famous
Golterman-tables~\cite{Go86}:
\begin{equation}
  O^{ik}(t) = \sum_{\vec{x}} s^{ik}_{\vec{x},t} \bar{\chi}_{\vec{x},t}
  \chi_{\vec{x},t}
\end{equation}
\begin{tabular}{|c|c|c|l|} \hline
  \rule[-2mm]{0cm}{6.5mm}$i$ 
  & $s^{ik}_x$ & $J^{PC}$ & particle \\*[1mm] \hline\hline
  1 
                &1&$\begin{array}{l}
                               0^{++}_s \\ 0^{-+}_a
                             \end{array}$ &
                            $\begin{array}{l}
                               \sigma \; {\rm (f}_0) \\ \pi^{(1)}
                             \end{array}$ \\ \hline
  2 
                &$\eta_{4x} \xi_{4x}$&
                            $\begin{array}{l}
                               0^{+-}_a \\ 0^{-+}_a
                             \end{array}$ &
                            $\begin{array}{l}
                               -  \\ \pi^{(2)}
                             \end{array}$ \\ \hline
  3 
                 &$\eta_{kx} \vep_x \xi_{kx}$&
                            $\begin{array}{l}
                               1^{++}_a \\ 1^{--}_a
                             \end{array}$ &
                            $\begin{array}{l}
                               a \\ \rho^{(3)}
                             \end{array}$ \\ \hline
  4 
                 &$\eta_{4x} \xi_{4x} \eta_{kx} \vep_x \xi_{kx}$&
                            $\begin{array}{l}
                               1^{+-}_a \\ 1^{--}_a
                             \end{array}$ &
                            $\begin{array}{l}
                               b \\ \rho^{(4)}
                             \end{array}$ \\ \hline
\end{tabular}\\[2mm]
The sign factors~$s^{ik}_x$ are composed of the standard
staggered phase factors~$\eta_{\mu x} = (-1)^{x_1+ \cdots +x_{\mu-1}}$,
$\xi_{\mu x} = (-1)^{x_{\mu+1}+ \cdots +x_4}$
and~$\vep_x = (-1)^{x_1 + \cdots + x_4}$.

The same continuum particles show up in different operators. Within large
error bars we did not observe up to now any contradiction to the flavour
symmetry restoration.

If one replaces the Higgs-sector of the standard model, the $\sigma$-meson
would be the candidate for a composite Higgs-boson. The $\pi$-meson would be
the Goldstone boson, which would later be eaten by the elektro-weak gauge bosons.

\subsection{Other Bosons}
We are further interested in the scalar and vector bosons, which
are also present in the~$U \phi$~sector without fermions.
The corresponding operators are
\begin{eqnarray}
  {\cal O}^{({\rm S})} (t) \hspace{-2mm}&=&\hspace{-2mm}
          \frac{1}{L^3} \sum_{\vec{x}} {\rm Re}
              \left\{\sum_{i=1}^{3} \phi^\dagger_{\vec{x},t} U_{(\vec{x},t),i}
                      \phi_{\vec{x}+\vec{i},t} \right\} ,
\label{OS} \\
  {\cal O}^{({\rm V})}_{i} (t) \hspace{-2mm}&=&\hspace{-2mm}
               \frac{1}{L^3} \sum_{\vec{x}} {\rm Im}
              \left\{ \phi^\dagger_{\vec{x},t} U_{(\vec{x},t),i}
                    \phi_{\vec{x}+\vec{i},t} \right\} \;, \\[-.5mm] \nonumber
                 & & \hspace{3cm}   i = 1,\,2,\,3 \;.
\label{OV}
\end{eqnarray}
Because of the fixed length of the scalar field, $\phi_x^\dagger\phi_x$ is a
trivial observable. Therefore one uses the suitable link products even in the
scalar case.

In principle we expect, that in the meson- and boson-operators with the same
quantum numbers also the same particles show up, but we couldn't confirm this
up to now.

\subsection{Theoretical Considerations}
For $\beta = 0$ the Lee-Shrock-transformation~\cite{LeShr87a} shows
that this model is equivalent to the Nambu-Jona-Lasinio model. It is known to
have a $\chi$PT (point {\bf N}) and dynamical mass generation in the chirally
broken phase. On the other hand this model is non renormalizable and a
continuum limit is not possible.

In the case $\kappa \rightarrow 0$ states with $\phi$, like the fermion F,
become infinitely heavy, whereas in the case $\kappa=\infty$ the fermion is
free and $m_{\rm F}=0$. It is very likely that $m_{\rm F}=0$ in the whole
chirally symmetric phase. Therefore it is of great interest to look for the
scaling near the $\chi$PT-line. On the right side of {\bf E}, where the phase
transition is first order, no scaling can be achieved. On the left hand side
of {\bf E} the question is open. We expect that the $\chi$PT of the NJL model
continues to $\beta>0$. So it may also be in the same universality class and
the model non renormalizable.  But at the point {\bf E} the universality class
probably changes, and thus the renormalizability might improve. Therefore our
greatest interest concerns the region near {\bf E}.

\section{Numerical results}
The calculations have been performed on $6^3\cdot16$ and $8^3\cdot24$
lattices for different $\beta$ and $\kappa$ near the endpoint {\bf E} and for some
intermediate $\beta$ on the {\bf NE} line.

\subsection{Fermions}
We observe in the whole chirally symmetric phase small fermion masses, which
can be linearly extrapolated in the bare fermion mass to very small values.
We expect that those small values are finite size effects and the values are
consistent with 0.

\begin{figure}[tbf]
{\centering
\fpsysize=7.5cm
\fpsbox[70 90 579 760]{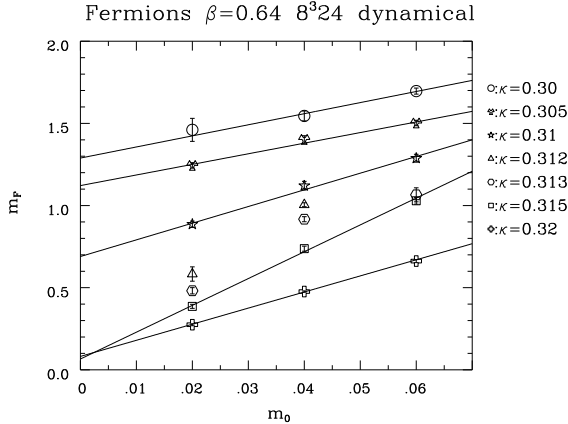}%
}%
\vspace*{-10mm}
\caption{The fermion mass as a function of the bare mass for different
  $\kappa$ at $\beta=.64$, which is approximately the value of the endpoint. The
  straight lines are fits to the data. The deviations from the linear behavior
  for intermediate $\kappa$ may be due to the phase transition.}
\label{fig:fer}
\vspace{-3mm}
\end{figure}
In the broken phase the masses increase very fast with decreasing $\kappa$,
the closer to the point {\bf E} the faster. In the first order regime the
fermion mass jumps to a nonzero value. In figure~\ref{fig:fer} the measured
fermion masses near the endpoint {\bf E} are shown.

\subsection{Bosons}
In contrast to fermions and mesons, the bosons have only little dependence on
the bare mass, but large finite size effects can be observed.

The mass of the scalar boson $m_{\rm S}$ shows a minimum at the point of the
phase transition. For large $\beta$ this is as a signal that chiral and Higgs
phase transitions coincide. For small $\beta$, where no Higgs phase transition
is present, this might be the effect of a cross over. The smaller $\beta$ is,
the larger the mass $m_{\rm S}$ and the less pronounced the minimum is.

We see a global minimum for the boson mass at $(\beta,\kappa) \approx
(.65,.31)$. Its value on the $6^3\cdot 16$ is 0.6(1) and on the $8^3\cdot 24$
lattice 0.4(1).  Measurements on larger lattices are required to show, whether
this mass vanishes on an infinite lattice, what might be expected.  The
coordinates of this point coincides within good precision with those of the
endpoint {\bf E} determined by the local observables.

\begin{figure}[tbf]
{\centering
\fpsysize=7.5cm
\fpsbox[70 90 579 760]{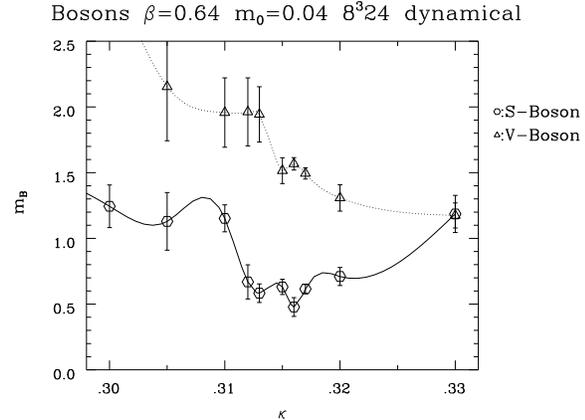}%
}%
\vspace*{-10mm}
\caption{The boson masses for fixed bare mass $m_0=.04$ at $\beta=.64$. The
  scalar mass (joined line) show a dip at the point of the PT. The vector
  boson (dotted line) doesn't scale at the phase transition.}
\label{fig:bos}
\vspace*{-1mm}
\end{figure}
Figure~\ref{fig:bos} shows scalar and vector boson mass for $\beta=.64$, where
we have most data. In this figure it can also be seen that the vector boson
doesn't scale.

\subsection{Mesons}
Until now we'd payed our main interest to $\sigma$- and $\pi$-meson. The
$\sigma$ is very hard to measure. We don't have enough statistics and
sufficiently good understanding of the first operator to present conclusive
results.

In the chirally broken phase the $\pi$-meson should behave like a Goldstone
boson. This can be checked by $(am_\pi)^2$ linearly going to
zero. In~\cite{LuFr95} we show a figure, which demonstrates this.

\section{Summary}
We have shown, that the model has in the broken chiral symmetry phase
essentially the expected properties. The results for the fermion and boson
masses look very promising. Also the $\pi$-meson seems to behave like a
Goldstone boson in the broken phase. The very interesting mass of the
$\sigma$-boson, which would be the composite Higgs, couldn't be determined up
to now.

The calculation have been performed on the HLRZ Cray Y-MP8/864 and the
`Landesvektorrechner' of NRW SNI/Fujitsu VPP 500.

\bibliographystyle{wunsnot}   



\end{document}